\documentclass[sigconf]{acmart}

\usepackage[utf8]{inputenc}
\usepackage{tikz}
\usetikzlibrary{
    positioning,    
    shapes.geometric, 
    arrows.meta,      
    fit, shapes,arrows,positioning,decorations.pathreplacing, arrows.meta, backgrounds
}

\copyrightyear{2026}
\acmYear{2026}
\setcopyright{cc}
\setcctype{by}
\acmConference[SVM '26]{4th International Workshop on Software Vulnerability Management }{April 12--18, 2026}{Rio de Janeiro, Brazil}
\acmBooktitle{4th International Workshop on Software Vulnerability Management (SVM '26), April 12--18, 2026, Rio de Janeiro, Brazil}
\acmPrice{}
\acmDOI{10.1145/3786165.3788442}
\acmISBN{979-8-4007-2397-1/2026/04}

\begin{document}

\title[An Approach to Generate Attack Graphs with a Case Study on Siemens PCS7 Blueprint]{An Approach to Generate Attack Graphs with a Case Study on Siemens PCS7 Blueprint for Water Treatment Plants}




\author{Lucas   Miranda}
\orcid{0009-0008-1589-0209}
\author{Carlos  Eduardo   Banjar}
\orcid{0000-0003-1576-3863}
\affiliation{%
  \institution{IC/UFRJ}
  \city{Rio de Janeiro}
  \country{Brazil}
}

\author{Daniel S. Menasche}
\orcid{0000-0002-8953-4003}
\affiliation{%
  \institution{PPGI/IC/UFRJ}
  \city{Rio de Janeiro}
  \country{Brazil}
}

\author{Anton Kocheturov}
\orcid{0000-0003-2549-9146}
\author{Gaurav Srivastava}
\orcid{0009-0005-2755-629X}
\affiliation{%
  \institution{Siemens}
  \city{Princeton}
  \country{USA}
}

\author{Tobias Limmer}
\orcid{0000-0001-8904-0620}
\affiliation{%
  \institution{Siemens}
  \city{M{\"u}nchen} 
  \country{Germany}
}

\renewcommand{\shortauthors}{Miranda, et al.}

\begin{abstract}
Assessing the security posture of Industrial Control Systems (ICS) is critical for protecting essential infrastructure. However, the complexity and scale of these environments make it challenging to identify and prioritize potential attack paths. This paper introduces a semi-automated approach for generating attack graphs in ICS environments to visualize and analyze multi-step attack scenarios. Our methodology integrates network topology information with vulnerability data to construct a model of the system. This model is then processed by a stateful traversal algorithm to identify potential exploit chains based on preconditions and consequences. We present a case study applying the proposed framework to the Siemens PCS7 Cybersecurity Blueprint for Water Treatment Plants. The results demonstrate the framework's ability to simulate different attack scenarios, including those originating from known CVEs and potential device misconfigurations. We show how a single point of failure can compromise network segmentation and how patching a critical vulnerability can protect an entire security zone, providing actionable insights for risk mitigation. 
\end{abstract}




\maketitle

\section{Introduction}
Industrial Control Systems (ICS) form the backbone of critical infrastructure, including water treatment plants, power grids, and manufacturing facilities~\cite{al2019a2g2v,ibrahim2023reinforcement}. The increasing connectivity of these systems to IT networks has exposed them to a growing number of cyber threats, making robust security assessment a necessity. However, manually analyzing complex ICS architectures to identify potential attack vectors is time-consuming and error-prone.

\begin{figure}[t]
    \centering
    \begin{tikzpicture}[
        node distance=0.45cm, 
        io/.style={
            rectangle, 
            draw, thick,
            fill=blue!5,
            align=center,
            font=\scriptsize,
            text width=0.62\columnwidth
        },
        process/.style={
            rectangle, 
            rounded corners=2pt,
            draw, thick,
            fill=green!5,
            align=center,
            text width=0.62\columnwidth,
            font=\scriptsize\bfseries,
            minimum height=2.6em
        },
        output/.style={
            ellipse, draw, thick,
            fill=red!5,
            align=center,
            text width=0.5\columnwidth,
            font=\scriptsize\bfseries
        },
        data/.style={
            font=\tiny\itshape,
            align=center,
            fill=white,
            inner sep=1pt
        },
        arrow/.style={
            -{Stealth[length=2.2mm, width=2.2mm]}, thick, draw=black!80
        }
    ]

    \node[font=\scriptsize\bfseries] (input_title) {Inputs:};
    \node[io, below=0.12cm of input_title] (blueprint) {\texttt{blueprint.json} \\ \tiny(Network Topology)};
    \node[io, below=of blueprint] (ssas) {\texttt{ssa.json} \\ \tiny(CVE Database)};
    \node[io, below=of ssas] (aliases) {\texttt{component\_aliases.json} \\ \tiny(Name Mapping)};

    \node[process, below=0.6cm of aliases] (phase1) {
        Phase 1: Topology Parser  \\
        \scriptsize Extracts nodes and edges from \texttt{blueprint.json}
    };

    \node[process, below=of phase1] (phase2) {
        Phase 2: Vulnerability Manager \\
        \scriptsize Maps CVEs to devices via aliases; converts CVSS/CPE to pre/post conditions.
    };

    \node[process, below=of phase2] (phase3) {
        Phase 3: Attack Graph Generator \\
        \scriptsize Stateful traversal computes valid exploitation paths.
    };

    \node[output, below=of phase3] (final_output) {
        Final Attack Graph \\ \tiny(Exploitation Paths)
    };

    \draw[arrow] (blueprint.west) -- ++(-0.35cm,0) |- (phase1.west);
    \draw[arrow] (ssas.east) -- ++(0.35cm,0) |- (phase2.east);
    \draw[arrow] (aliases.east) -- ++(0.55cm,0) |- (phase2.east);

    \draw[arrow] (phase1) -- node[right=0.06cm, data] {Unenriched\\Topology Graph} (phase2);
    \draw[arrow] (phase2) -- node[right=0.06cm, data] {Enriched Graph\\(with vulnerabilities)} (phase3);

    \draw[arrow] (phase3) -- (final_output);

    \end{tikzpicture}
    \caption{Workflow: inputs $\rightarrow$ processing   $\rightarrow$   attack graph.}
    \label{fig:methodology-flow-compact} 
        \Description{Flowchart of the proposed pipeline. Three input files (blueprint.json for network topology, ssa.json for a CVE database, and component\_aliases.json for name mapping) feed three processing phases: Topology Parser, Vulnerability Manager, and Attack Graph Generator. Arrows indicate intermediate artifacts (an unenriched topology graph and an enriched graph with vulnerabilities) leading to the final output, an attack graph of exploitation paths.}
\vspace{-0.2in}
\end{figure}

Attack graphs are a powerful formalism for modeling and analyzing multi-step attacks, providing a clear visualization of how an attacker can chain vulnerabilities to exploit critical assets. While the concept is not new, its application to complex ICS environments, coupled with automation, remains a significant challenge.

This paper presents a semi-automated approach to generate and analyze attack graphs for ICS networks. Our solution consists of a modular framework that processes network architecture data and vulnerability information to simulate attack scenarios. This allows security analysts to proactively identify weaknesses, evaluate the impact of potential threats, and prioritize mitigation efforts.

Our main contributions are:
\begin{itemize}
    \item A semi-automated, multi-stage methodology for generating attack graphs in ICS environments, combining a network topology parser, a vulnerability manager, and a graph-based simulation engine~\cite{miranda2025svmAttackGraphSim}.
    \item A case study demonstrating the practical application of the proposed framework  on the Siemens PCS7 Cybersecurity Blueprint for Water Treatment Plants \cite{Siemens_WWTP_SecureGuideline_2023}, a representative model for modern industrial facilities. We show how  the framework provides actionable insights for improving the system's security posture.
\end{itemize}

\textbf{Outline. }  
Section~\ref{sec:methodology} introduces our semi-automated attack graph generation methodology and 
Section~\ref{sec:case-study} presents a case study on the Siemens PCS7 Cybersecurity Blueprint.  
Section~\ref{sec:discussion} discusses broader implications, including integrating quantitative exploitation estimation. 
Section~\ref{sec:related-work} reviews related research and Section~\ref{sec:conclusion} concludes. 

\section{Methodology} \label{sec:methodology}

 Our approach is implemented in a software prototype composed of three main components: a \emph{Topology Parser}, a \emph{Vulnerability Manager}, and an \emph{Attack Graph Generator}. 
As illustrated in Figure~\ref{fig:methodology-flow-compact}, the process begins with the Topology Parser, which ingests a structured network description and produces a directed connectivity graph. 
This graph is then enriched by the Vulnerability Manager, which matches assets to CVEs (Common Vulnerabilities and Exposure) and transforms each vulnerability into pre- and post-conditions. 
Finally, the Attack Graph Generator performs a stateful traversal of the enriched graph to enumerate feasible multi-step attack paths and output an interactive visualization. 



\subsection{Topology Parser and Vulnerability Manager}
The first stage involves parsing a representation of the ICS network architecture. This is based on a structured JSON file that defines all network components (e.g., firewalls, switches, PLCs, servers), their interconnections, and their grouping into security zones. This component constructs a directed graph where nodes represent components and buses, and edges represent the connections between them. This foundational graph serves as the map for the subsequent vulnerability and attack analysis.

Once the network topology is established, the Vulnerability Manager enriches this model with security-relevant information. This process is twofold.

First, the component matches devices from the topology graph to known vulnerabilities. Using a manually curated list of aliases (e.g., product families, alternative names) for each device, it searches a database of vendor advisories (e.g., Siemens Security Advisories, SSAs) to find relevant CVEs. In its current state, the framework does not account for specific firmware versions or applied patches, which can lead to the inclusion of CVEs that might have been mitigated in newer releases. However, the decision to not filter by patch level is grounded in the difficulty of patch management within industrial environments, where legacy software is common and the application of patches is often delayed due to operational uptime requirements~\cite{10.1145/3084455}.

The second phase of this process converts each discovered CVE into a set of preconditions (also referred to as precedents, that define what an attacker needs to exploit the
vulnerability) and consequences~\cite{Ou2005MulVAL,islam2008heuristic,barrere2021analysing} (also referred to as postconditions~\cite{10.1145/3176258.3176339, 10.1145/3297280.3297401}, that  define what an attacker gains after a successful exploit). To automate this translation, we implemented a \emph{rule-based approach} that analyzes vulnerability attributes sourced from the \emph{NVD}, such as CVSS vector strings, and the presence of specific keywords within CVE descriptions. This methodology adopts the rules derived from prior studies \cite{10.1145/3176258.3176339, 10.1145/3297280.3297401}, whose implementations are publicly available in our repository~\cite{miranda2025svmAttackGraphSim}.

In cases where a CVE does not match any predefined rule in our knowledge base, 
we adopt a conservative approach by assigning it a precondition of \texttt{OS(ADMIN)} (requiring full privileges already) 
and a postcondition of \texttt{NONE}, representing no change in system state. This approach reduces the likelihood of generating false-positive attack paths. In practice, this treatment is equivalent to excluding such vulnerabilities from the analysis.

The complete classified vulnerability landscape is summarized in Table \ref{tab:aggregate_summary}. The most striking finding from the precondition data in Table \ref{tab:aggregate_summary} is the number of vulnerabilities that require no prior access. We found 2052 classified vulnerabilities with a precondition of \texttt{NONE}. 
This represents a security gap, as these flaws are directly accessible to attackers without the need for initial credential compromise.

On the impact side, the data clearly points to the most severe outcome. We identified 803 vulnerabilities that directly result in \texttt{OS(ADMIN)} privileges. This means a successful exploit grants the attacker full administrative control over the operating system. 
For the majority of other vulnerabilities (2726) we were unable to match any predefined rule, and we conservatively set their consequence as \texttt{NONE}, as discussed  above.


\begin{table}[h]
\centering
\caption{Preconditions and Consequences}
\Description{Table summarizing counts of classified vulnerabilities by required precondition (APP or OS privileges, or none) and by consequence (privilege gained or none).}
\label{tab:aggregate_summary}
\begin{tabular}{p{5cm} r}
\toprule
\textbf{Category} & \textbf{Count} \\
\midrule
\multicolumn{2}{c}{\textbf{Preconditions}} \\
APP (ADMIN) & 150 \\
APP (USER)  & 356 \\
NONE        & 2052 \\
OS (ADMIN)  & 402 \\
OS (USER)  & 598 \\
\midrule
\multicolumn{2}{c}{\textbf{Consequences}} \\
NONE        & 2726 \\
OS (ADMIN)  & 803 \\
OS (USER)   & 29 \\
\bottomrule
\end{tabular} \vspace{-0.2in}
\end{table}

\subsection{Attack Graph Generator}
The core of our methodology is the Attack Graph Generator. This component uses a stateful graph traversal algorithm to simulate an attacker's movement through the network, where each node in the graph represents a specific network device. The simulation begins from a user-defined starting node.

The algorithm iteratively explores all possible attack paths as follows:
\begin{enumerate}
    \item \textbf{Local Privilege Escalation:} First, the algorithm checks if the attacker can exploit any vulnerabilities on their \textit{current node} to gain higher privileges. The attacker's state is updated accordingly; notably, any successful device exploitation at this stage, regardless of the privilege level attained, establishes a foothold for further action.
    \item \textbf{Lateral Movement:} Next, the algorithm identifies all reachable target nodes connected to the current node.
    \item \textbf{Exploitation Check:} For each reachable target, the algorithm compares the attacker's current privilege state (gained from step 1) against the preconditions of all vulnerabilities on that target node.
    \item \textbf{Exploitation:} If the attacker's state satisfies the preconditions of a vulnerability, the attacker exploits the vulnerability on the target node. The attacker's state is then updated based on the consequence of this new exploit, and the process repeats from this new node.
\end{enumerate}

If a node (e.g., a firewall) has no vulnerabilities whose preconditions can be met, it acts as a barrier, and the algorithm stops traversing that path. If it is vulnerable, it becomes a bridge to the next network zone. The final output is a directed graph of all identified multi-step attack paths.

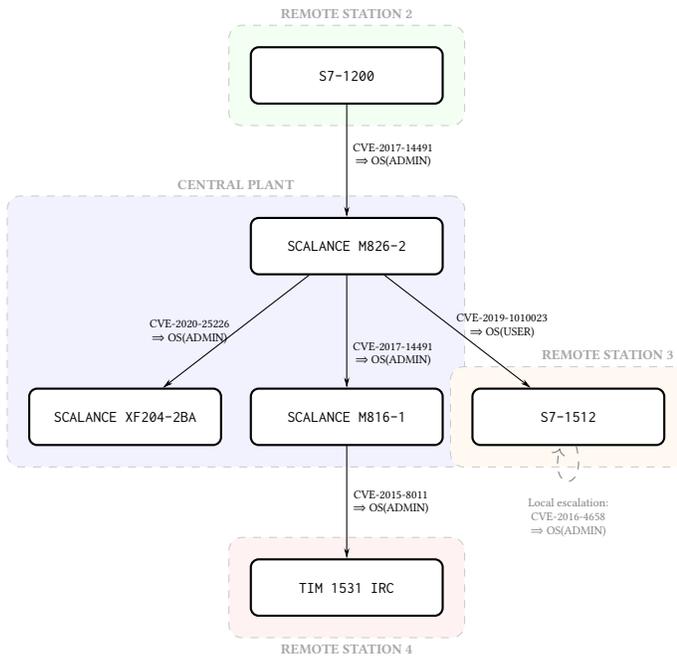
\begin{figure}[ht]
\centering
\begin{tikzpicture}[
    scale=0.75,
    transform shape,
    device/.style={rectangle, rounded corners=3pt, draw=black, thick, minimum width=34mm, minimum height=10mm, align=center, font=\small, fill=white},
    arrow/.style={-{Stealth[length=4pt, width=2pt]}, thin},
    dashedarrow/.style={-{Stealth[length=4pt, width=2pt]}, thin, dashed, gray},
    lab/.style={font=\scriptsize, align=center},
    zone_box/.style={draw=gray!40, dashed, rounded corners=5pt, inner sep=8pt},
    zone_label/.style={font=\footnotesize\bfseries, gray!70}
]

\node[device] (S7_1200) {\texttt{S7-1200}};
\node[device, below=2cm of S7_1200] (M826) {\texttt{SCALANCE M826-2}};

\node[device, below left=2cm and 0.5cm of M826] (XF204) {\texttt{SCALANCE XF204-2BA}};
\node[device, below=2cm of M826] (M816) {\texttt{SCALANCE M816-1}};
\node[device, below right=2cm and 0.5cm of M826] (S7_1512) {\texttt{S7-1512}};

\node[device, below=2cm of M816] (TIM) {\texttt{TIM 1531 IRC}};

\begin{scope}[on background layer]
    \node[zone_box, fill=blue!5, fit=(M826) (XF204) (M816), label={[zone_label]above:CENTRAL PLANT}] (CP) {};
    
    \node[zone_box, fill=green!5, fit=(S7_1200), label={[zone_label]above:REMOTE STATION 2}] (RS2) {};
    
    \node[zone_box, fill=orange!5, fit=(S7_1512), label={[zone_label, xshift=7mm]above:REMOTE STATION 3}] (RS3) {};
    
    \node[zone_box, fill=red!5, fit=(TIM), label={[zone_label]below:REMOTE STATION 4}] (RS4) {};
\end{scope}

\draw[arrow] (S7_1200) -- node[lab, right, pos=0.45] {CVE-2017-14491\\$\Rightarrow$ OS(ADMIN)} (M826);

\draw[arrow] (M826) -- node[lab, left, pos=0.5] {CVE-2020-25226\\$\Rightarrow$ OS(ADMIN)} (XF204);

\draw[arrow] (M826) -- node[lab, right, pos=0.45] {CVE-2019-1010023\\$\Rightarrow$ OS(USER)} (S7_1512);

\draw[arrow] (M826) -- node[lab, right, pos=0.7] {CVE-2017-14491\\$\Rightarrow$ OS(ADMIN)} (M816);

\draw[arrow] (M816) -- node[lab, right, pos=0.5] {CVE-2015-8011\\$\Rightarrow$ OS(ADMIN)} (TIM);

\draw[dashedarrow] (S7_1512) edge[loop below] node[lab, below, yshift=-2mm] {Local escalation:\\CVE-2016-4658\\$\Rightarrow$ OS(ADMIN)} ();

\end{tikzpicture}
\caption{Attack graph fragment with   exploit paths and one local escalation} 
\Description{Diagram grouping devices into four zones (Remote Station 2, Central Plant, Remote Station 3, Remote Station 4). Arrows labeled with CVE identifiers show exploit-driven transitions from S7-1200 to SCALANCE M826-2, then to SCALANCE XF204-2BA, SCALANCE M816-1, S7-1512, and finally TIM 1531 IRC. A dashed self-loop at S7-1512 indicates a local escalation.}
\vspace{-0.3in}
\label{fig:attackgraph}
\end{figure}

  Figure \ref{fig:attackgraph} depicts an \emph{inter-zone attack}. \texttt{SCALANCE M826-2} and \texttt{SCALANCE M816-1} are at the Central Plant Zone. The initial foothold at \texttt{S7-1200}, at Remote Station 2 Zone, leads to the exploitation of \texttt{SCALANCE M826-2}, followed by lateral exploitation and a local escalation on \texttt{S7-1512}, at Remote Station 3 Zone; the \texttt{TIM 1531 IRC} device is located at Remote Station 4 Zone. Each directed edge represents a successful vulnerability-driven transition under the algorithm's propagation rules. The figure therefore illustrates pivoting and escalation across zones (Remote Station 2 → 3 → 4).

CVE-2017-14491 affects the SCALANCE M-800 family, which is why it appears at two points in the graph. CVE-2020-25226 targets the SCALANCE X-200 switch family, while CVE-2019-1010023 impacts the SIMATIC S7-1500 CPU family. Finally, CVE-2015-8011 affects the TIM 1531 IRC module.

\section{Case Study: Siemens PCS7 Blueprint} \label{sec:case-study}

The Siemens PCS7 Cybersecurity Blueprint for Water Treatment Plants adopts a defense-in-depth architecture that segments the network into four main zones: the \emph{Enterprise} or IT network, the \emph{Demilitarized Zone (DMZ)} for intermediary services, the \emph{Building Control} zone hosting remote PLCs, and the \emph{Central Plant} zone containing the main process controllers. 
Two industrial firewalls enforce this segmentation: the \emph{Front Firewall}, which separates the Enterprise network from the DMZ, and the \emph{Back Firewall}, which isolates the DMZ from the control layers.

Figure~\ref{fig:attackgraph} illustrates a less common attack scenario that begins with the exploitation of a PLC (\texttt{S7-1200}) at the top of the figure.
Although PLCs are usually internal assets, this scenario models a compromised remote station device acting as the initial foothold, showing how an adversary could escalate privileges upward through interconnected field devices.  
In this section, we focus on the more typical case in which the attack originates from the network perimeter and attempts to move inward across zones.

Simulations of the baseline PCS7 configuration capture proper segmentation: exploiting a network-facing vulnerability on the \texttt{Front Firewall} allowed compromise of the DMZ but not the Building or Central Plant zones.  
When a misconfiguration was introduced in the \texttt{Back Firewall} the attacker was able to pivot from the DMZ through the \texttt{SCALANCE M826-2} gateway into the Building zone, reaching the \texttt{S7-1512} PLCs.  
A similar weakness in a \texttt{SCALANCE XF204-2BA} switch further enabled access to the Central Plant zone, demonstrating how a single inter-zone misconfiguration can collapse intended isolation boundaries.

Conversely, removing the vulnerability from the \texttt{Front Firewall} broke the entire attack chain, leaving no feasible path from external networks to process-control assets.  
This shows that keeping perimeter firewalls and gateway devices patched yields a large defensive benefit, effectively safeguarding downstream zones within the PCS7 architecture.

While these analyses provide a qualitative view of how multi-step attacks can propagate across zones, from the network perimeter toward process-control assets, they also motivate a need for quantitative assessment. 
This transition enables a data-driven comparison of defensive priorities, linking the qualitative topology, e.g., as shown in Figure~\ref{fig:attackgraph}, to the quantitative metrics reported in Table~\ref{tab:target_probabilities_nodes}.

\section{Likelihood Estimation of Exploitation} \label{sec:discussion}

We propose extending our framework by integrating Reliability Block Diagrams (RBDs), used to model how individual component failures impact overall system availability. This analogy directly maps the concept of a \emph{component failure} in reliability engineering to a \emph{component exploitation} within our attack graph. Consequently, a \emph{system failure} in an RBD model corresponds to the successful \emph{exploitation of a crown jewel} asset.

This mapping allows us to model attack scenarios using the two primary RBD configurations. From the \textit{attacker's} perspective, a single multi-step attack path identified by our Attack Graph Generator is analogous to a \emph{series configuration}. In this model, the attacker must successfully exploit every vulnerability in the chain to reach the target; a failure at any single step breaks that entire path. Conversely, from the \emph{defender's} standpoint, the complete set of all distinct attack paths leading to a critical asset represents a \emph{parallel configuration}. This models the redundancy of attack vectors: the asset is exploited if \textit{any single path} is successful. Therefore, a successful defense requires that \textit{all} potential attack paths are mitigated.

We take the final Attack Graph as input to this RBD analysis~\cite{mell2024measuring}, and use the Exploit Prediction Scoring System (EPSS) \cite{10.1145/3436242} to parametrize the RBDs. The EPSS provides an empirically-derived probability that a known vulnerability will be exploited in the wild \footnote{Given that EPSS scores are updated daily \cite{Ravalico2024EPSS}, the scores used in this study correspond to a snapshot taken on October 27, 2025.}. Alternatively, other studies have opted to use CVSS-based metrics to estimate exploitation probabilities \cite{10838376}.

The probability of an attack through a  single path \( j \) succeeding, denoted by \( P_{\text{path}_j} \), is modeled as a \emph{series system}. Therefore, it is given by the product of the individual exploit success probabilities \( P_{\text{EPSS}_i} \) for all \( i \) exploits along that path. In turn, the total probability of compromising the target, \( P_{\text{target}} \), is modeled as a \emph{parallel system} and corresponds to one minus the probability that all \( j \) paths fail:
\begin{align}
P_{\text{path}_j} &= \prod_{i \in \text{path}_j} P_{\text{EPSS}_i}, \\
P_{\text{target}} &= 1 - \prod_{j \in \text{paths}} \left(1 - P_{\text{path}_j}\right).
\end{align}

We analyzed the likelihood of exploitation for each target in the network based on the enumerated attack paths and EPSS scores of associated vulnerabilities. For each target, all other devices in the network are considered potential sources of attack. The results are reported in Table~\ref{tab:target_probabilities_nodes} and indicate that several SCALANCE devices, acting as network pivots, exhibit near-certain probability of exploitation due to multiple high-probability paths. PLCs such as S7-1512 and S7-1510 show substantial probability, whereas other devices, including management and monitoring hosts, have low or negligible probability. This occurs because there are fewer paths leading to these devices, or the vulnerabilities present in the existing paths are significantly more difficult to exploit.

\begin{table}[h!]
\centering \vspace{-0.1in}
\caption{Estimated success probability ($P_{target}$) accounting for all feasible paths from all sources.}
\Description{Table listing each target device, its estimated compromise probability, number of feasible attack paths, and summary statistics (mean, median, max) of path length.}
\label{tab:target_probabilities_nodes}
\footnotesize
\begin{tabular}{
    l
    r 
    r 
    |r 
    r 
    r
}
\toprule
\textbf{ } & 
\textbf{} & 
\textbf{ } &
\multicolumn{3}{c}{\textbf{Path length}} \\
\textbf{Target} & 
\textbf{$P_{target}$} & 
\textbf{Paths} &
\textbf{Mean} & 
\textbf{Median} &
\textbf{Max} \\
\midrule
SCALANCE M826-2 & 0.999990 & 15 & 2.13 & 2.00 & 3 \\
SCALANCE M876-4 & 0.999990 & 15 & 2.13 & 2.00 & 3 \\
SCALANCE M816-1 & 0.999983 & 15 & 2.20 & 2.00 & 3 \\
Energy Manager Pro & 0.988465 & 13 & 2.00 & 2.00 & 3 \\
S7-1512 & 0.811917 & 15 & 3.00 & 3.00 & 4 \\
S7-1510 & 0.811917 & 15 & 3.00 & 3.00 & 4 \\
TIM 1531 IRC & 0.281317 & 15 & 3.07 & 3.00 & 4 \\
SCALANCE XF204-2BA & 0.094664 & 15 & 2.27 & 2.00 & 3 \\
SCALANCE XF204-2BA DNA & 0.094664 & 15 & 2.27 & 2.00 & 3 \\
PCS 7 Mgt/SINEC NMS Control & 0.016789 & 13 & 1.92 & 2.00 & 2 \\
SINEC NMS Operation & 0.016789 & 13 & 1.92 & 2.00 & 2 \\
Anomaly Detection Center & 0.000000 & 1 & 1.00 & 1.00 & 1 \\
\bottomrule
\end{tabular} \vspace{-0.1in}
\end{table}

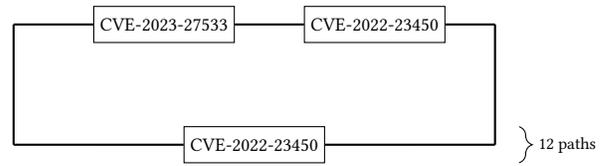
\begin{figure}[t]
\centering
\begin{tikzpicture}[
  scale=0.8, transform shape,
  circuit/.style={draw, rectangle, minimum width=2.2cm, minimum height=0.6cm, align=center},
  line/.style={thick},
  lab/.style={font=\small, align=center}
]

  \node[circuit] (A1) at (2.5,0) {CVE-2023-27533};
  \node[circuit] (A2) at (6,0)   {CVE-2022-23450};
  \draw[line] (0,0) -- (A1.west);
  \draw[line] (A1.east) -- (A2.west);
  \draw[line] (A2.east) -- (8,0);

  \node[circuit] (C1) at (4,-2.0) {CVE-2022-23450};
  \draw[line] (0,-2.0) -- (C1.west);
  \draw[line] (C1.east) -- (8,-2.0);

  \draw[decorate, decoration={brace, amplitude=5pt}] (8.4,-1.7) -- (8.4,-2.3)
    node[midway, right=6pt, lab] {12 paths};

  \draw[line] (0,0) -- (0,-2.0);
  \draw[line] (8,0) -- (8,-2.0);

\end{tikzpicture}
\caption{All possible attack paths leading to the Energy Manager Pro device. 
Of the 13 paths shown, 12 require only one exploitation step.}
\Description{Reliability-block style diagram showing multiple parallel attack paths toward the Energy Manager Pro target. One path contains two CVEs in series (CVE-2023-27533 then CVE-2022-23450). A second repeated branch shows a single-step path with CVE-2022-23450, annotated as repeated 12 times.}
\label{fig:attackgraph1}
\end{figure}

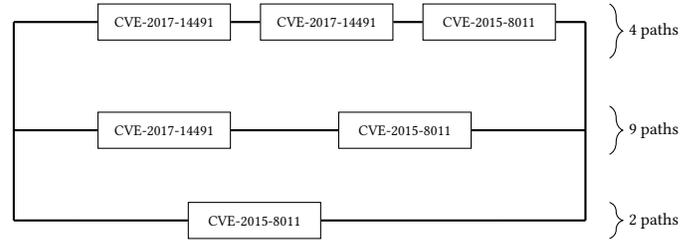
\begin{figure}[t]
\centering
\begin{tikzpicture}[
  scale=0.8, transform shape,
  circuit/.style={draw, rectangle, minimum width=2.2cm, minimum height=0.6cm, align=center, font=\footnotesize},
  line/.style={line width=0.8pt},
  lab/.style={font=\small, align=center}
]

  \node[circuit] (A1a) at (2.5,0) {CVE-2017-14491};
  \node[circuit] (A1b) at (5.2,0) {CVE-2017-14491};
  \node[circuit] (A1c) at (7.9,0) {CVE-2015-8011};
  \draw[line] (0,0) -- (A1a.west);
  \draw[line] (A1a.east) -- (A1b.west);
  \draw[line] (A1b.east) -- (A1c.west);
  \draw[line] (A1c.east) -- (9.5,0);

  \draw[decorate, decoration={brace, amplitude=5pt}] (9.9,0.3) -- (9.9,-0.6)
    node[midway, right=6pt, lab] {4 paths};

  \node[circuit] (B1a) at (2.5,-1.8) {CVE-2017-14491};
  \node[circuit] (B1b) at (6.5,-1.8) {CVE-2015-8011};
  \draw[line] (0,-1.8) -- (B1a.west);
  \draw[line] (B1a.east) -- (B1b.west);
  \draw[line] (B1b.east) -- (9.5,-1.8);

  \draw[decorate, decoration={brace, amplitude=5pt}] (9.9,-1.4) -- (9.9,-2.2)
    node[midway, right=6pt, lab] {9 paths};

  \node[circuit] (C1) at (4,-3.3) {CVE-2015-8011};
  \draw[line] (0,-3.3) -- (C1.west);
  \draw[line] (C1.east) -- (9.5,-3.3);

  \draw[decorate, decoration={brace, amplitude=5pt}] (9.9,-3.0) -- (9.9,-3.6)
    node[midway, right=6pt, lab] {2 paths};

  \draw[line] (0,0) -- (0,-3.3);
  \draw[line] (9.5,0) -- (9.5,-3.3);

\end{tikzpicture}
\caption{All possible attack paths leading to the TIM 1531 IRC. Of the 15 paths shown, nine consist of two CVEs, four consist of three CVEs, and two consist of a single CVE.} 
\Description{Reliability-block style diagram with parallel branches representing attack paths toward the TIM 1531 IRC target. Some branches chain CVE-2017-14491 and CVE-2015-8011 (with repetition annotations), and some branches contain only CVE-2015-8011 as a single-step path.}
\vspace{-0.2in}
\label{fig:attackgraph2}
\end{figure}

We can observe from Table~\ref{tab:target_probabilities_nodes} that the number of possible attack paths is not, by itself, a determinant factor for the value of \(P_{\text{target}}\). 
Two cases illustrate this behavior well: the \textit{Energy Manager Pro} and the \textit{TIM 1531 IRC}. 
The former has fewer paths leading to it (13 in total), meaning that it can be reached from fewer initial points; 
however, it still presents a higher overall probability than the latter. 
As shown in Figure~\ref{fig:attackgraph1}, the paths toward the \textit{Energy Manager Pro} are generally shorter, 
requiring the exploitation of fewer CVEs. 
Conversely, the paths in Figure~\ref{fig:attackgraph2} that lead to the \textit{TIM 1531 IRC} are longer, 
making it less probable for an attacker to successfully traverse any given path.

Another important factor influencing the discrepancy between the \(P_{\text{target}}\) values is the EPSS of the vulnerability present in the target device itself. 
Since each attack chain necessarily ends with the exploitation of the target’s own CVE, 
the EPSS value of that vulnerability is incorporated into all paths that reach the target. 
In this case, \( \text{EPSS}(\text{CVE-2022-23450}) = 0.33344 \) and 
\( \text{EPSS}(\text{CVE-2015-8011}) = 0.04146 \). 
Because the latter is considerably lower, the paths ending with CVE-2015-8011 --- corresponding to the 
\textit{TIM 1531 IRC} --- become even less likely to be successfully exploited, 
further explaining the lower \(P_{\text{target}}\) observed for that device.

\section{Discussion and Related Work}

Understanding how adversaries can exploit vulnerabilities through multi-step campaigns has been a long-standing research problem, leading to a rich literature on attack graph generation, vulnerability modeling, and quantitative reliability assessment. 
This section reviews key developments in these areas, with particular attention to their application in ICS and SCADA environments.

\textbf{Foundations of Attack Graphs. }
Automated generation of attack graphs is a well-established research area. 
Early work by Sheyner et al.~\cite{Sheyner2002} formalized the use of model checking to generate and analyze multi-step attack scenarios, showing how reachability and privilege escalation could be automatically derived from network configurations. 
Rule-based systems such as MulVAL~\cite{Ou2005MulVAL} later provided a scalable way to reason about host configurations and vulnerabilities, becoming a reference point for automated analysis. 
Subsequent works introduced probabilistic reasoning~\cite{Sawilla2008}, Bayesian inference~\cite{Frigault2008}, and abstraction techniques to handle large graphs.

\textbf{CVSS and Vulnerability Modeling. }
Many attack graph frameworks rely on standardized vulnerability data to define exploit preconditions and consequences. 
The CVSS~\cite{CVSSv3} provides structured exploitability attributes, which can be mapped to logical constraints in attack graphs~\cite{Cheng2012, Gallon2015}. 
Works such as~\cite{Gallon2015,Cheng2012} explicitly use CVSS fields to determine exploit feasibility and attack path length, while others integrate temporal or environmental metrics to refine prioritization.

\textbf{ICS and SCADA-Specific Attack Graphs. }
While traditional attack graphs focus on IT networks, ICS and SCADA environments introduce unique challenges such as vendor-specific devices, legacy protocols, and segmented network architectures. 
Prior work has extended attack graphs to model ICS-specific semantics and process-level consequences~\cite{Patapanchala2016, Barrere2020, operational}. 
For example, PLCs and firewalls are treated as first-class components with different roles and exposure levels, which affects how attack paths propagate across zones. 
However,   ICS-specific approaches   require significant manual modeling.

\textbf{Firewall Semantics and Misconfiguration. }
Firewall and access control modeling has also been addressed in prior work. 
MulVAL and related systems typically model firewalls either as binary barriers between zones or as fine-grained access control lists (ACLs) when configuration data are available~\cite{Ou2005MulVAL,Cheng2012}. 
Several studies~\cite{Durumeric2014,wool2004quantitative} emphasize that misconfigurations often play as significant a role as CVEs in real attacks, highlighting the need to support “what-if” analysis in security assessment.
Indeed, 
a key feature of our approach is the ability to model risks beyond documented CVEs. While patch management for known vulnerabilities is crucial, many real-world security incidents stem from misconfigurations (e.g., weak passwords, improper firewall rules) or zero-day exploits. By allowing the creation of ``artificial vulnerabilities,'' our framework enables security analysts to conduct ``what-if'' analyses. 

\textbf{Exploitation. }
Recent works have proposed combining attack graphs with probabilistic models to estimate the likelihood of system exploitation. 
Approaches such as EPSS~\cite{EPSS2021} provide data-driven probabilities that a CVE will be exploited in the wild. 
These have been combined with RBD analogies to compute path and asset exploitation probabilities~\cite{Cheng2012, Frigault2008}. 
While these approaches are known, integrating EPSS, CVSS-based modeling, and ICS semantics in a single automated pipeline remains underexplored.

\textbf{Our Work. }
Our work builds on these lines of research by:
(i) explicitly mapping CVSS attributes to logical preconditions and consequences used in automated attack graph traversal;
(ii) incorporating ICS-specific semantics (e.g., PLCs, SCALANCE switches, firewalls) with zone-based or ACL-based firewall modeling; and 
(iii) integrating probabilistic exploitation estimation using EPSS in combination with attack graph enumeration.

\label{sec:related-work}

\section{Conclusion and Future Work} \label{sec:conclusion}

We presented a method to generate attack graphs for Industrial Control Systems by integrating network topology and vulnerability data. The Siemens PCS7 case study showed that the approach effectively identifies critical failure points and supports remediation analysis.  Future work includes refining the attack simulator with realistic privilege escalation and lateral movement, refining the product matching process to account for specific software versions and available patches, and explicitly modeling firewall configurations to capture concrete misconfigurations. Additionally, we plan to leverage a digital twin to replicate identified attack paths.

\begin{acks}
This work was   supported by CAPES, FAPERJ   grant E-26/204.268/2024, and   CNPq grants 444956/2024-7  and    315106/2023-9. 
\end{acks}

\bibliographystyle{ACM-Reference-Format}
\bibliography{sample-base}

\end{document}